\documentclass{rmaa}

\usepackage{psfrag,color}
\usepackage[latin1]{inputenc}

\usepackage{url,hyperref,lineno,microtype,subcaption}
\usepackage[onehalfspacing]{setspace}
\usepackage{graphicx}	
\usepackage{amsmath}	
\usepackage{bm}		
\usepackage{color}

\usepackage{natbib}

\usepackage{ulem}

\def\apjl{ApJ Letters}
\def\aap{A\&A}

\title{An alternative approach to the Finger of God in large scale structures}

\author{
  Luis Salas,\altaffilmark{1} 
  and Irene Cruz-Gonz\'alez\altaffilmark{2}}

\altaffiltext{1}{Instituto de Astronom\'ia, Universidad Nacional Aut\'onoma de M\'exico, Ensenada, B.~C., M\'exico.}

\altaffiltext{2}{Instituto de Astronom\'ia, Universidad Nacional Aut\'onoma de M\'exico, Ciudad de M\'exico, M\'exico.}

\shortauthor{Salas, \& Cruz-Gonz\'alez}
\shorttitle{Alternative approach to the Finger of God}

\fulladdresses{
\item Irene Cruz-Gonz{\'a}lez: Instituto de Astronom{\'i}a, Universidad Nacional Aut{\'o}noma de M{\'e}xico, Circuito Exterior, Ciudad Universitaria, Ciudad de M{\'e}xico 04510 (irene@astro.unam.mx).
\item Luis Salas: Instituto de Astronom{\'i}a, Universidad Nacional Aut{\'o}noma de M{\'e}xico, 
M{\'e}xico, Apdo. Postal 877, Ensenada B.~C.  22810, M{\'e}xico
  (salas@astro.unam.mx).}

\listofauthors{L. Salas, \& I. Cruz-Gonz{\'a}lez}
\indexauthor{Salas, L.}
\indexauthor{Cruz-Gonz{\'a}lez, I.}

\abstract{ 

It is generally accepted that linear theory of growth of structure
under gravity produces a squashed structure in the two-point
correlation function (2PCF) along the line of
sight (LoS). On the other hand, the observed radial spread out structure known
as Finger of God (FoG)  is attributed to non-linear effects.  
In this paper we argue that the squashed structure associated with
the redshift-space ($s-$) linear theory 2PCF is obtained only when this function is 
displayed in real-space ($r-$), or when the mapping from $r-$ to $s-$space is approximated.
We find a way to solve for the mapping function $\bm s(\bm r)$ 
that allows us to display the $s-$space 2PCF 
properly in a grid in $s-$space, by using plane of the sky projections of the
$r-$ and $s-$ 2PCFs. We show that even in the simplest case of the linear Kaiser spectrum  with a conservative power-law $r-$space 2PCF, a structure quite similar to the FoG is observed in the small scale
region, while in the large scale the expected squashed structure 
is obtained. This structure depends on only three parameters.

}

\resumen{

Com\'unmente se acepta que la teor\'ia de colapso 
lineal gravitacional produce una estructura comprimida a lo largo de la 
visual en la funci\'on de correlaci\'on de dos puntos (2PCF). Por otro 
lado, la estructura conocida como {\it Finger of 
God} (FoG) se ha atribuido a efectos no-lineales.
Aca argumentamos que la estructura asociada con  el espacio de 
corrimiento al rojo ($s-$) de la 2PCF de la teor\'ia lineal, s\'olo 
se obtiene  cuando esta funci\'on es desplegada en el espacio-real ($r-$) 
\'o cuando el mapeo de  $r-$ al $s-$ se calcula mediante una aproximaci\'on. 
Encontramos una forma de resolver para la funci\'on de mapeo $\bm s(\bm r)$ 
que nos permite visualizar correctamente la  $s-$  2PCF en una
malla en $s-$,  utilizando proyecciones en el plano del cielo para 
ambas 2PCFs, $r-$ y $s-$. 
Mostramos que a\'un en el caso mas simple, el de un espectro de Kaiser con
ley de potencia para la 2PCF del $r-$, es posible apreciar a peque\~na escala una 
estructura similar a FoG, mientras que a gran escala se obtiene la 
estructura comprimida esperada. Dicha estructura solo depende de tres par\'ametros.
}

\addkeyword{galaxies: clusters: general}
\addkeyword{galaxies: quasars: general}
\addkeyword{cosmology: theory}
\addkeyword{cosmology: large scale structure}

\begin{document}
\maketitle

\section{Introduction}

A spherical object observed at a distance in its longitudinal ($||$) 
and transversal ($\bot$) dimensions,
should provide a test of different cosmological models, as first proposed 
by \citet{ap79}.
The Alcock-Paczy{\'n}ski parameter,  hereafter  $AP$, basically the ratio of $||$ 
to $\bot$ dimensions, takes a value of one at 
redshift zero, and increases with $z$ with a strong dependence on the value 
of the cosmological 
parameters that make up the Hubble function, introducing   a cosmological
distortion  to the large scale structure observations.  This apparently simple 
comparison is, however, 
greatly complicated by several factors.  First, real-space measurements are 
not directly attainable and one has to rely on redshift-space.  Then, if the proposed 
object consists of a cluster of quasars or galaxies, or a statistical ensemble of such,  
proper motions of its constituents, either derived from gravitational collapse 
or virialized conditions,  distort redshift-space measurements causing a degeneracy problem \citep[e.g.,][]{hamilton98}.
On cosmological scales, clusters of galaxies or quasars are among the 
most simple geometric structures that one may conceive.  Even if single 
clusters may have  non-spherical or filamentary structures, those should
be randomly oriented.  As we probe more distant
clusters, observations become biased towards brighter and widely 
separated members, and the numbers become statistically insignificant.   
A superposition of many such  clusters may reduce
the problem   while  retaining spherical symmetry. 
The two-point correlation function (2PCF), and its Fourier transform, the power spectrum, have been fundamental tools in these studies for more than 40 years \citep[e.g.,][]{peebles80}.

Overdense clusters or associations separate from the Hubble flow due to their
own gravity, which results in peculiar velocities of its members that distort
redshift-space observations.  When gravitational 
fields are small, velocities are well described by linear theory of gravitational collapse \citep{peebles80}. 
In the study of these clusters, the 2PCF was initially conceived as a
single entity $\xi$ that could be evaluated in either  real ($r-$) or redshift ($s-$) space \citep{peebles80}.  \citet{davis_peebles_83} even mentioned that when observing
the local universe, if the peculiar velocities were small by comparison, $s-$space would directly reproduce $r-$space and one would have $\xi(\bm r) = \xi (\bm s)$.  That should be the case for distant objects, although one should be careful not to mix up the notions of distant from each other and distant from the observer.  
In the case of the  CfA Redshift Survey \citep[e.g.,][]{1983ApJS...52...89H}, as described in \citet{davis_peebles_83} 
peculiar velocities were significant, and the authors chose to go from real $\xi(r)$ to observable $\xi(s)$ by means of a convolution with a pair-wise velocity distribution, 
tailored to  approach the Hubble flow at large distances, known as the {\it streaming model}.  The convolution integral would at the same
time convert $r-$space to $s-$space coordinates.  However, the same function
$\xi$ would be obtained as a result of the convolution of $\xi$ with a function of velocity, which constitutes an inconsistency.
Later on, \citet{kaiser87}, hereafter K87, showed that gravitationally induced peculiar velocities by
gravitational collapse of overdense structures in the linear regime, produces a power spectrum $P^{(s)}$ for $s-$space
different from the one  $P^{(r)}$ for $r-$space, that is 
two different functions for the power spectrum.  Both are,
however, functions of the $r-$space Fourier frequency {$\bm k$}.  Then, while $P^{(r)}({\bm k})$ is a spherically symmetric function, $P^{(s)}({\bm k})$ shows an elongation
along the line of sight (LoS) direction.  Later on, \citet{hamilton92} translated these results to configuration space obtaining the 2PCF in its
two flavors:  $\xi^{(r)}({\bm r})$ and  $\xi^{(s)}({\bm r})$.  Again $\xi^{(r)}({\bm r})$
is symmetric and the possibility of a power-law $r^{-\gamma}$ is considered, as had been
historically accepted \citep[e.g.,][who favored $\gamma = 1.8$ ]{peebles80}.  
Also in perfect agreement with K87,  $\xi^{(s)}({\bm r})$ shows a
squashing along the LoS direction.  \citet{hamilton98} presents in great detail the assumptions that led to his results.  He starts by defining 
selection functions $n^{(r)}(r)$ and $n^{(s)}(s)$ for $r-$space  and $s-$space and by numerical conservation
obtains a complicated high order expression (his eq. 4.28) for the density contrast $\delta^{(s)}$. 
From that one can obtain
the 2PCF, but a series of approximations are needed (the linear case) first 
to reduce the right hand side of the equation and end up in his eq. 4.30 for $\delta^{(s)}({\bm s})$.  Then, he performs one extra
assumption, $\delta^{(s)}({\bm r}) = \delta^{(s)}({\bm s})$, which is not justified 
by the linear approximation. This changes the left hand side of 
the equation directly to $\delta^{(s)}({\bm r})$.  It may be argued that
this approximation is valid in the distant case mentioned above.  
Consequently,  one could easily write  $\xi^{(s)}({\bm s})$ in place
of  $\xi^{(s)}({\bm r})$, shifting between one form and the other as needed.  That 
is an imperative because observable 2PCF are inevitably obtained in $s-$space. 

Since then many authors have tried the Kaiser linear approximation facing this dilemma 
and have performed  similar  approximations.  
In the description of 2PCF in redshift-space, due to the multipole expansion   of
the inverse Lagrangian operator derived from the corresponding power spectrum in Fourier space \citep{hamilton92}, there appears a dependence with $\mu$, the cosine of the angle between the    ${\bm r}$  (real space) vector and the LoS: 
 $ ~ \mu{(\bm r)}\,=\,r_{||}/|{\bm r}|  $.
However, it has been a common practice to approximate  $\mu$  from redshift-space coordinates as either $ \mu{(\bm s)} = s_{||}/|{\bm s}|$ or  $\mu{(\bm {cs})} = c_{||} s_{||}/\sqrt{ c_{\bot}^2 s_{\bot}^2 + c_{||}^2 s_{||}^2 }$
\citep[e.g.,][]{ms96, Nakamura_1998_ApJ_494_13, lc14}.  
Yet in some other cases the approximation $r_{||} = s_{||}$ is specifically made \citep[e.g.,,][]{2006MNRAS.368...85T} calling it
the ``distant observer" approximation.  But as mentioned above this is really meant to mean a wide separation
approximation and does not apply in the small scale regime. Furthermore, the ``distant observer"  name is
also used for the plane-parallel case \citep[e.g.,][]{percival2008}, adding to confusion. 
In some other cases the substitution $r_{||} = s_{||}$ is just performed with no further comment \citep[e.g.,][]{hawkins2003}.
Another facet of the same problem has been to expand the redshift-space correlation function 
as a series of harmonics of that same $\mu{(\bm s)}$, rather than the actual $\mu({\bm r})$ derived in linear theory 
\citep[e.g.,][]{guo2015, chuang2012, marulli2017}.  While this
is certainly a valid approach, the conclusions of linear theory, like the existence of only monopole, quadrupole  and  hexadecapole terms in the Legendre  polynomial expansion, are not really applicable to the $\mu{(\bm s)}$ case.
All these forms of the approximation are really one and the same, and to avoid further confusion 
(like the term ``distant observer") 
we decided to call it the $\mu(\bm s)$ approximation.

When observational data is used to construct the 2PCF $\xi^{(s)}(\bm s)$, it is generally
true that  simple linear theory predictions are not kept. On one hand, the predicted compression along
the viewing direction is observed, but as one approaches the LoS axis 
the observed structure is mostly dominated by an elongated feature \citep[e.g.,][]{hamaus2015}, 
usually called Finger of God \citep{huchra88},   hereafter FoG. 
Prominent examples of FoG were found in the Coma Cluster by \citet{deLapparent86}  
and in the Perseus cluster by \citet{Wegner93}.  The FoG feature is also 
commonly observed in 
the 2PCF of statistical aggregates \citep[e.g.,][]{hawkins2003},
making it a  common feature in large scale structure.

Many studies have been conducted  to explain this discrepancy.  In general
non-linear processes are invoked.  Sometimes the non-linearities are assigned to
virial relaxation in the inner regions of clusters, while others explore the
non-linear terms in the approximation in the derivation of the K87 result.
In these categories, we mention a small
sample of representative literature.  
Kinematic relaxation, 
like the virialized motion of cluster members in the inner regions 
\citep{kaiser87,hamaus2015}, are explored 
by introducing a distribution of pair-wise peculiar velocities for cluster components.
There are at least two ways of doing so: First, the {\it streaming model} where a velocity
distribution $f(V)$ is convolved with $\xi^{(r)}(\bm r)$ to obtain $\xi^{(s)}(\bm s)$, without
using the K87 result, similar to \citet{davis_peebles_83} but differentiating 
$\xi^{(s)}$ from  $\xi^{(r)}$. 
More recent work on distribution functions 
take great care on this issue \citep{2011JCAP...11..039S,2012JCAP...11..014O,2012JCAP...02..010O}
by directly obtaining  the power spectra in redshift-space as a function of the $s-$space wave-number. Unfortunately,
the expression that results for the power spectra is rather complicated, even
when it is conveniently expressed as a series in mass weighted velocity moments.
However, it is possible to obtain FoG
structures in $\xi^{(s)}(\bm s)$ maps by the convolution with simple velocity distributions, at the same
time that a mapping from $r-$ to $s-$space takes place  \citep[e.g.,][]{ 2004PhRvD..70h3007S}.
Paradoxically, it is not that easy to obtain the traditional peanut-shape structure that is generally recognized as the K87 limit in  $\xi^{(s)}(\bm s)$, unless the limit
$s \sim r$ is once again invoked. 
Second, in the phenomenological {\it dispersion model} \citep[c.f.,][]{2004PhRvD..70h3007S, 2006MNRAS.368...85T}
a linear K87 spectrum is multiplied  in Fourier space by a velocity distribution.  This can be seen
as a convolution in configuration space, as in   \citet{hawkins2003}, but the procedure has the disadvantage
that it obtains the same function $\xi^{(s)}$ as the result of the convolution of $\xi^{(s)}$ and
$f(V)$.   It has to be noted, however, that very good fits to the observed data are obtained
by this procedure.  The same is true for the fits to numerical simulation results at mid spatial frequencies
obtained by similar procedures in e.g., \citet{marulli2017}. 
In the streaming model, the velocity distribution function can also be obtained from the
interaction of galaxies with dark mater halos \citep[e.g.,][] {2006MNRAS.368...85T,2007MNRAS.374..477T},
via the halo occupation distribution formalism.

Apart from kinematics,  non-linear terms also arise in the expansion of the
mass conservation or continuity equation in $r-$ and $s-$spaces to obtain the power spectrum or the
2PCF \citep[e.g.,][]{2008PhRvD..77f3530M, 2010PhRvD..82f3522T,2016JCAP...08..050Z}.  
Preserving only first order terms yields the K87 result.  However, a full
treatment of all the terms is possible with the use of perturbative
methods. There are diverse techniques: standard, Lagrangian, re-normalized, resumed Lagrangian
\citep[for a comparison see][]{percival2008,2011MNRAS.417.1913R}.  The latter authors however, conclude
that the failure of these methods  to fit the 
$l$=2 and 4 terms in the expansion  $\xi^{(s)}_l(r)$ on quasi-linear scales of 30 to 80 $h^{-1}$ Mpc,  must be due to inaccuracies in the mapping between $r$- and $s$-spaces.  So, they favor again the {\it streaming model}.  Clearly, there is
still substantial debate on this subject.

In most of these works the necessity to translate their results to observable 2PCFs, $\xi^{(s)}(\bm s)$, is not really
addressed.   Most authors prefer
to display their results in Fourier space as $s$-space power spectrum $P^{(s)}(\bm k_r)$   \citep[e.g.,][]{2008PhRvD..77f3530M,2012JCAP...11..014O}, but with $k_r$ in $r$-space; or 
display its moments $P^{(s)}_l(\bm k_r)$ 
\citep[e.g.,][]{2010PhRvD..82f3522T,2016JCAP...08..050Z};
or power spectra with $k_s$ in $s-$space $P^{(s)}(\bm k_s , \mu_s)$ \citep[e.g.,][]{2012JCAP...02..010O}.
Other authors display the correlation function in $r$-space, either as $\xi^{(r)}(\bm r)$
\citep[e.g.,][]{2008PhRvD..77f3530M}
or $\xi^{(s)}(\bm r)$ \citep[e.g.,][]{2007MNRAS.374..477T,2011MNRAS.417.1913R, 2012JCAP...11..014O}, or its moments $\xi^{(s)}_l(s)$ 
\citep[e.g.,][]{2010PhRvD..82f3522T}
for $l$=2. Few works try to display directly the 2PCFs $\xi^{(s)}(\bm s)$ 
\citep[e.g.,][]{ms96, Nakamura_1998_ApJ_494_13,2006MNRAS.368...85T, lc14},
but as already mentioned above, usually perform the $\mu(\bm s)$ approximation that amounts to really obtaining $\xi^{(s)}(\bm r)$ instead.

To further complicate matters,  redshift-space distortions are often treated
separately from the cosmological distortions.  Both are not easily discernible 
because both produce stretching or
squashing in the LoS direction \citep{hamilton98,hamaus2015}.
This degeneracy could in
principle be resolved because the cosmological and peculiar velocity signals
evolve differently with redshift, but in practice the uncertain {evolution of bias (the dimensionless growth rate for visible matter, see eq.~\ref{eq:dpmn26})} complicates the problem \citep{Ballinger96}.  Furthermore,
\citet{kaiser87} and \citet{hamilton92,hamilton98} do not consider cosmological 
distortions in their analysis of peculiar motions. Since the earlier works, the inclusion of cosmological distortions has been attempted by several authors
\citep[e.g.,][]{ms96, hamaus2015}.

In this paper, we show that a structure quite similar to FoG can be obtained in $\xi^{(s)}(\bm s)$ directly in the
linear theory limit of K87.  That is, without invoking virial relaxation or the 
{\it streaming model}, nor the non-linearities studied in perturbation theory, but 
just by avoiding the $\mu(\bm s)$ approximation, in any of its forms ($\mu=s_{||}/|\bm s| $, ``distant observer" or  $r_{||} = s_{||}$ ), the FoG structure is recovered. This will be accomplished by solving for the function $\bm r ( \bm s)$ with the aid
of the projected correlation function of both 2PCFs : $\xi^{(s)}(\bm s)$  and $\xi^{(r)}(\bm r)$.  We will stay on the academic power-law approximation $\xi^{(r)}(\bm r) \sim r^{-\gamma}$ in order to be able to show a closed form for the result, and to prove the main point of this paper, i.e. that the FoG feature is derived in the simplest case.

{We start with a detailed definition of $r-$ and $s-$space, noting that }
frequently $s-$space  is expressed in distance units as is $r-$space.
But in doing so, one multiplies by a scale factor that invariably introduces
a cosmological parameter in the definition; and as a result the named
$s-$space is no longer purely observational.
{Later on the factor is solved by introducing a fiducial cosmology and
solving for the real values.}
An example can be seen in the analysis made by \citet{padwhite2008} in Fourier 
space and \citet{xu2013} in configuration space. The latter recognize the need 
of introducing a two-step
transformation, one isotropic dilation and one warping transformation, 
to transform from real fiducial to
real space. However, the real fiducial space is actually 
redshift-space, and this identification is missing in these works.

Therefore, we argue (c.f., Section~\ref{sec:sec2}) that it is convenient to define 
the observable-redshift-space ${\bm \sigma}$ (${\sigma}$-space) given by the simple redshift differences and subtended
angles that are truly observable, and that do not depend on any choice of 
cosmological parameters.
Multiplying by a units function (scale factor) produces the physical 
redshift-space ($s-$space): ${\bm s} = K({\bm \Omega},z) ~ {\bm \sigma}$,
that is isomorphic to the observable   ${ \sigma}$-space, but has actual 
distance units that 
are dependent on a particular cosmological set of parameters {\bm $\Omega$} 
and the redshift $z$.  The
$K({\bm \Omega},z)$ function is chosen so that the physical redshift-space 
is related to real space  ${\bm r}$  by a unitary Jacobian
independent of redshift. So that no additional scaling  is needed,
and the only remaining difference will be precisely in shape.
That is why   ${\bm \sigma}$ and  ${\bm s}$  
are more alike, and thus can both be named redshift-space; ${\bm \sigma}$ is the
observable redshift-space while  ${\bm s}$ is the physical redshift-space. Then, 
the transformation to real-space necessarily 
goes through redshift distortions.

Furthermore, when we introduce peculiar non-relativistic 
velocities in this scheme, 
we will show that it is possible
to keep the same relation between observable and physical redshift-spaces, 
${\bm s}$ and ${\bm \sigma}$,
and that the \citet{kaiser87} effect is recovered independently of 
redshift (see Section~\ref{sec:sec3}).  
That is, 
now redshift-space will also show an additional gravitational
distortion with respect to real-space. 

To solve for the relation between real-space and redshift-space, we will rely on projected correlations.  Projections of the 2PCF in the plane of the sky have been widely used to avoid the complications of dealing with
unknown components in redshift-space \citep[e.g.,][]{davis_peebles_83}.  This has the advantage that in the case of a symmetric 2PCF in real-space, the 3-D structure can be inferred from the projection.  
We will show in Section~\ref{sec:sec4} that since the projections of the 
2PCF in real-space and in redshift-space are bound to give the same profile, 
a relationship can be obtained for the real-space coordinate $r_{||}$ as
a function of the corresponding one in redshift-space $s_{||}$.  From this, 
we solve for $\mu{(\bm r)}$ in real-space, and show that a different view of the
redshift-space 2PCF emerges. The main result is that the redshift-space 2PCF
presents a distortion in the LoS direction which looks similar to the 
ubiquitous FoG.
This is due to a strong anisotropy that arises purely from linear theory  
and   produces   a change in scale as one moves into the on-axis LoS 
direction.   
As we move out of the LoS, a structure somewhat more squashed than the
traditional result by the $\mu(\bm s)$ approximation is obtained. 
As this effect has been missed before (to the best of our knowledge), 
we provide a detailed derivation in Sections~\ref{sec:sec2} to \ref{sec:sec4}, and show examples of the derived 2PCFs in redshift-space (Section~\ref{sec:sec5}). Finally, in Section~\ref{sec:sec6} we summarize our main conclusions.

\section{redshift-space}\label{sec:sec2}

Consider the Friedmann-Lema\^itre-Robertson-Walker metric \citep[e.g.,][]{harrison93} written in units of distance and time as follows:
\begin{equation}\label{eq: dpmn1}
{ds}^2=c^2{dt}^2 - d r^2=c^2{dt}^2 - a(t)^2\left({d\chi }^2+S_k(\chi
)^2({d\theta }^2+\sin ^2(\theta ){d\varphi }^2)\right)  ,
\end{equation}
with \  $S_k=(~sin ~, ~ {Identity}~,~ sinh~ )$ for $k=(1,0,-1)$.
Then the co-moving { present-time }   length of an object $d r^0$  that is observed longitudinally is 
related to a variation in the observed redshift $d z$ by
\begin{equation} \label{eq:dpmn2}  
d r_{||}^0 = \frac{c d z} {H(z) },
\end{equation}
where $H(z)$ is the Hubble function and the $^0$ superindex is used to define the present time $t_0$. Similarly, an object with a transversal
co-moving dimension ${d r}_{\bot}^0$
subtends an angle $d \theta$ 
given by the angular co-moving distance \citep[e.g.,][]{hogg99} as 
\begin{equation}  \label{eq:dpmn3} 
\frac {{d r}_{\bot}^0}{{d \theta}}=a_0S_k\left(\frac c{a_0}\int _0^z\frac{{dz}'}{H(z')}\right),
\end{equation}
where $a_0$ is the present day scaling parameter of the metric.

Observationally one measures redshift differences $dz$ and subtended 
angles $d\theta$. We then 
define the observable redshift-space adimensional quantities ($d\sigma_{||}~, d\sigma_{\bot}$) as
\begin{equation}  \label{eq:dpmn4} 
d\sigma_{||} = dz 
\end{equation} and
\begin{equation}  \label{eq:dpmn5} 
d\sigma_{\bot} = z  d\theta.
\end{equation}
The physical 
redshift-space sizes $ds_{||}$ and $ ds_{\bot}$ can then be defined in terms of ${\bm \sigma}$ as
\begin{equation}  \label{eq:dpmn6} 
ds_{||} = K({\bm \Omega},z) d\sigma_{||}
\end{equation} and
\begin{equation}  \label{eq:dpmn7} 
ds_{\bot}= K({\bm \Omega},z) d\sigma_{\bot}, 
\end{equation}
where $K({\bm \Omega},z)$ has distance units and depends on the cosmology, 
represented here symbolically by the ${\bm \Omega}$ terms. 
The relation between real-space and physical redshift-space is then obtained 
from eqs. \eqref{eq:dpmn2} to \eqref{eq:dpmn7}, that is:
\begin{equation}  \label{eq:dpmn8} 
dr^0_{||} = c_{||} ds_{||}
\end{equation} and
\begin{equation}  \label{eq:dpmn9} 
dr^0_{\bot} = c_{\bot} ds_{\bot},
\end{equation}
with
\begin{equation}  \label{eq:dpmn10} 
c_{||}=\frac{ c } {K({\bm \Omega},z) ~ H(z) }
\end{equation} and
\begin{equation}  \label{eq:dpmn11} 
c_{\bot} = \frac{   a_0 }{z ~ K({\bm \Omega},z)  } ~ S_k\left(\frac c{a_0}\int _0^z\frac{{dz}'}{H(z')}\right).
\end{equation}

It is clear then that  the \citet{ap79} 
function $AP(z)$, that tests redshift distortions of a particular 
cosmology, can be written as
\begin{equation}  \label{eq:dpmn12} 
 AP(z) =  \frac{c_{\bot} (z)} {c_{||}(z)}  =\frac{a_0} {c} ~\frac{H(z)} z ~ S_k\left(\frac c{a_0}\int _0^z\frac{{dz}'}{H(z')}\right). 
\end{equation}
Furthermore, from the transformation of physical redshift-space 
with coordinates $(ds_{\bot},ds_{\bot},ds_{||})$ into
real-space $(dr_{\bot}^0,dr_{\bot}^0,dr_{||}^0)$ we get a Jacobian 
\begin{equation} \label{eq:dpmn13}  
\left| {\frac  { d^3 \bm s}  { d^3 \bm r} } \right| = \frac{1}{c_{||}(z)} ~ \frac{1}{c_{\bot}^{2}(z)} .
\end{equation}
In order for this transformation to preserve scale we need a
unitary Jacobian. This condition can be achieved simply by the following condition:
\begin{equation} \label{eq:dpmn14}  
K({\bm \Omega},z)= \frac {c}{H(z)} ~ AP(z)^{2/3}, 
\end{equation}
as can be seen from eqs. \eqref{eq:dpmn10} to \eqref{eq:dpmn12}. 
Here the dependence on the cosmology is made explicit through the Hubble function.
Note that the resulting scale factor $K({\bm \Omega},z) $ approaches the Hubble radius $a_H = c/H_0$ as $z \rightarrow 0$ and decreases 
approximately as $1/(1+z)$ thereafter. Also note that for redshift $z > 0$, the physical scale that transforms all dimensions of redshift-space, 
contracts isotropically.  Also we remark that $c_{||} $ and $ c_{\bot}$
are of order unity as $z \rightarrow 0$, and satisfy $c_{\bot}/c_{||} = AP(z)$ for all $z$.  In fact we have \citep[see also][]{xu2013}:
\begin{equation} \label{eq:dpmn15}   
c_{||}(z) = AP(z)^{-2/3},
\end{equation}
and
\begin{equation} \label{eq:dpmn16}   
c_{\bot}(z)  = AP(z)^{1/3}.
\end{equation}
 
Peculiar velocities modify the observed redshift, and therefore alter the relation
between real-space and redshift-space giving rise to kinematic distortions. Suppose the
near-end of an object is at rest at redshift $z$, while the far-end is moving with
peculiar non-relativistic velocity $\vec{\rm v }$.  Then it will appear Doppler
shifted to an observer at rest at the far-end position, causing eq.~\eqref{eq:dpmn2} to get the form
\citep[see also][]{ms96,hamaus2015}: 
\begin{equation} \label{eq:dpmn17}
 c dz = H(z)~ dr_{||}^0 + (1+z) ~ (\vec{\rm v} \cdot \hat{r}),
\end{equation}
where $\hat{r}$ points in the direction of the far-end, at an angle $d\theta$ from the near-end.
Since $ \vec{\rm v} \cdot \hat{r} = {\rm v}_{||} + {\rm v}_{\bot} ~d \theta$, 
then for small  angular 
separations ($d\theta << 1$) the perpendicular component of the peculiar velocity may be 
trivialized. Therefore eq.~\eqref{eq:dpmn8} gets modified to 
\begin{equation} \label{eq:dpmn18}  
dr_{||}^0  = c_{||} ~ ( ds_{||} -  ds_{\rm v} ).
\end{equation}
where $ds_{\rm v}$ (in physical redshift-space) is given by
\begin{equation} \label{eq:dpmn19}  
ds_{\rm v} = K({\bm \Omega},z) ~d\sigma_{\rm v} 
\end{equation}
and $d\sigma_v$ (in observable redshift-space) is given by
\begin{equation} \label{eq:dpmn20}  
d\sigma_{\rm v} = (1+z) ~ \frac{{\rm v}_{||}}c.
\end{equation}
And through the similarity of eqs. \eqref{eq:dpmn19} and \eqref{eq:dpmn20} 
with eqs. \eqref{eq:dpmn6} and \eqref{eq:dpmn4}, we note that the concepts 
of observable redshift-space and physical redshift-space can be extended to 
include peculiar motions as well.

\section{Two point correlation function}\label{sec:sec3}

Let  ${\bm r}$  be real-space Euclidean co-moving coordinates in the close vicinity of a point at redshift $z$, what was defined as $dr$ in the
previous section.  Then for azimuthal symmetry around the line of sight (aligned to the third  axis) we have   ${\bm r} $    = $ (  dr_{\bot}^0,  dr_{\bot}^0,dr_{||}^0)$. Let  ${\bm s}$  denote physical redshift-space coordinates around the same point (in the same tangent subspace), with the third axis along the line of sight.  Then, from eqs.  \eqref{eq:dpmn9} and \eqref{eq:dpmn18},  the Jacobian is 
\begin{equation} \label{eq:dpmn21}  
 \left| {\frac  { d^3 \bm s} { d^3 \bm r} } \right|  = \frac{1}{c_{||}(z)~ c_{\bot}^{2}(z)} ~ \left(  1 + {\frac {(1+z)}   {H(z)}} ~~
{\frac {\partial {\rm v}_{||} } {\partial r_{||}}} \right) =  1 + {\frac {(1+z)}  {H(z)}} ~~
{\frac{\partial {\rm v}_{||} }   {\partial r_{||}}},
\end{equation}
where we have used the unitary condition on eq. \eqref{eq:dpmn13} to eliminate the $c_{||}(z) ~ c_{\bot}^{2}(z)$ term.
In going from  ${\bm r}$  to   ${\bm s}$  space,  the density change can be related to the  change in volume $V$, and the Jacobian by the equation
\begin{equation}\label{eq:dpmn22}  
\left( {\frac {d\rho}   \rho}\right)_{\bm s-r} = - {\frac{dV} V} = 1 - \left| { \frac { d^3 \bm s} { d^3 \bm r} } \right|.
\end{equation}
This can also be expressed in terms of the contrast density ratios in $\bm s$ and $\bm r$ spaces defined such that
\begin{equation} \label{eq:dpmn23} 
\left( {\frac {d\rho} \rho}\right)_{\bm s-r} = \delta^{(s)} - \delta^{(r)},
\end{equation}
where $\delta^{(s)}$ and $\delta^{(r)}$ are two distinct scalar functions of position in either space.
This particular definition of $\delta^{(s)}$  requires knowledge of
the real-space  selection function \citep{hamilton98}, which makes it rarely a first choice. However, the procedure given below  allows us precisely to solve for the function ${\bm r}({\bm s})$.

In linear theory, \citet[][]{peebles80} shows in eqs. 14.2 and 14.8  that an overdensity of mass $\delta(\bm{r})$ creates a peculiar velocity field similar to the acceleration field produced by a mass distribution.  As such, it can be derived from a potential  function whose Laplacian is the overdensity itself \citep[e.g.,][]{marion2004}  times a constant which is time (or redshift) dependent.  That is
\begin{equation} \label{eq:dpmn24}  
{\bf {\rm v}}({\bm r} ) =  -{\frac{ H(z)~f(z)}{ (1+z)}} ~{\bm \nabla} \nabla^{-2} \delta^{(r)}_m({\bm r}),
\end{equation}
where $\nabla$ is the gradient and $\nabla^{-2}$ is inverse Laplacian, and
\begin{equation}\label{eq:dpmn25}
f(z) = {\frac{a(z)}{D(z)}}{ \frac{dD } {da}}.
\end{equation}
Here $D(z)$ is the growth factor, the temporal component of density. Note that  in \citet{peebles80} coordinates are given in the expanding background model  ${\bm x}$  which relate to present time real-space coordinates by ${\bm r } = a_0 {\bm x}$; this brings about the $(1+z)$ factor to eq.~\eqref{eq:dpmn24}. The $m$ subscript to $\delta$ emphasizes that all mass is responsible for the velocity field,  while $\delta$ without the subscript refers to visible mass  in the form of galaxies or quasars.  To account for the difference, it is customary to introduce a bias factor $b(z)$ and define the dimensionless growth rate for visible   matter 
\begin{equation}\label{eq:dpmn26}
\beta(z) = {\frac {f(z)}{b(z)}}.
\end{equation} 
Then from eqs. \eqref{eq:dpmn21} to \eqref{eq:dpmn24} we get:
\begin{equation} \label{eq:dpmn27} 
\delta^{(s)}({\bm r}) =  \left(1+   \beta(z) ~ \partial_{||}^2 ~ \nabla^{-2} \right)\delta^{(r)}({\bm r}), 
\end{equation}
where $\partial_{||}$ denotes $\partial / \partial r_{||}$ in real space.  Note that if we had not required 
a unity Jacobian (c.f., eq.~\ref{eq:dpmn13}), then eqs.~\eqref{eq:dpmn21} and \eqref{eq:dpmn22} would not had canceled
out the  $1 - c_{||}(z)^{-1}c_{\bot}(z)^{-2}$ term.  We note that this term is not small when $K(\Omega,z) $ is a constant, and will vary by one order of magnitude
as $z \rightarrow 1$,  and up to three orders of magnitude as $z \rightarrow 10$. So the transformation between observable and physical redshift-spaces cannot be neglected \citep[contrary to][assumption]{ms96}.

The square modulus of the Fourier transform of eq. \eqref{eq:dpmn27}  gives an expression for the power spectrum, or the Fourier transform of the autocorrelation function (2PCF) $\xi$, which generalizes \citet{kaiser87} results for any
redshift $z$
\begin{equation}\label{eq:dpmn28}  
\widetilde{\xi^{(s)}}({\bm k}) =  \left(1+   \beta(z) ~ \mu_k^2 \right)^2 
\widetilde{\xi^{(r)}}({\bm k}),    
\end{equation}
where $\mu_k = k_{r3}/|{\bm k_r} |$    is the cosine of the angle between the $k_{r3}$ component and the wave number vector  ${\bm k_r}$  in real-space; and it arises by the Fourier transform property of changing differentials
into products. Note that wave number vectors in real-space also differ from their counterparts in redshift-space by the unknown velocity field in eq. \eqref{eq:dpmn17}. 

Fourier transforming back into coordinate space gives \citet{hamilton92} result:
 
\begin{equation}\label{eq:dpmn29}   
  \xi^{(s)}({\bm r}) =  \left(1+   \beta(z) ~  \partial_{||}^2 ~\nabla^{-2} \right)^2\xi^{(r)}({\bm r}) .
\end{equation}
Note that this equation is written in a way that all terms in the right hand side are 
real-space coordinates   ${\bm r}$   dependent, as is the case for the derivatives and 
inverse Laplacian. 
Recalling that the solution of the Laplace equation in spherical
coordinates consists of spherical harmonics in the angular coordinates and a power series
in the radial part, one can write for the case of azimuthal symmetry
\begin{equation} \label{eq:dpmn30}
\xi^{(s)}({\bm r}) = \sum_{l=0} \xi_l(r) ~ P_l(\mu{(\bm r)})
\end{equation}
where $P_l(\mu{(\bm r)})$ are the Legendre polynomial,
\begin{equation}\label{eq:dpmn31}
{\mu{(\bm r)}} =  {\frac {r_{||}} {|{\bm r}|}},
\end{equation}
explicitly defined for real-space coordinates, and the harmonics are given by the coefficients $\xi_l(r)$  that can be obtained from eq.\eqref{eq:dpmn30}  through orthogonality properties as
\begin{equation} \label{eq:dpmn32}
\xi_l(r) = {\frac{(2l+1)} 2} \int_{-1}^1 P_l(\mu{(\bm r)}) ~\xi^{(s)}({\bm r}) ~d\mu{(\bm r)}.
\end{equation}
Substituting  eq.~\eqref{eq:dpmn29} in \eqref{eq:dpmn32} for the case of spherical symmetry in real-space ($\xi^{(r)}({\bm r}) = \xi^{(r)}(r) $), one gets  by direct evaluation the classical result given by
\citet{hamilton92}, see also \citet{hawkins2003}.  That result consists of 
only three terms, monopole,  quadrupole  and  hexadecapole 
($ l = 0, 2, 4$), all the others evaluate to zero. It is important to note that this is not true when the expansion of eq.~\eqref{eq:dpmn30} has been done in $\mu{(\bm s)}$ as is assumed by several authors \citep[e.g.,][]{guo2015, chuang2012, marulli2017}.

When the 2PCF could be approximated by a power-law, $ \xi^{(r)}(r) = (r/r_0)^{-\gamma}$, the solution for 
eq.~\eqref{eq:dpmn29} can be written as
\begin{equation} \label{eq:dpmn33}
\xi^{(s)}({\bm r}) = g(\gamma,\beta,\mu{(\bm r)}) ~ \xi^{(r)}(r).
\end{equation} 
where $g(\gamma,\beta,\mu{(\bm r)})$ has been written in several equivalent forms \citep{hamilton92, ms96, hawkins2003}. One of these is the following
\begin{equation} \label{eq:dpmn34}
g(\gamma,\beta,\mu{(\bm r)}) = 1 + 2 {\frac {1-\gamma \, \mu(\bm r)^2}  {3-\gamma}} ~ \beta(z)  +{ \frac {  \gamma (\gamma+2) \, \mu(\bm r)^4  -  6 \gamma \, \mu(\bm r)^2 + 3} { (3-\gamma) (5-\gamma) }} ~ \beta(z)^2.
\end{equation}
This function takes values greater than 1 for the equatorial region  ($\mu{(\bm r)} \rightarrow 0$), and less than 1 for the polar axis ($\mu{(\bm r)} \rightarrow 1$). 
Alternatively, it has been mentioned that the quadrupolar term 
in the multipole expansion dominates the hexadecapole. As a result of either argument
the 2PCF { $\xi^{(\bm s)} (\bm r) $ } seems squashed { with a peanut shape when displayed in $r$-space }, in agreement with common knowledge. 

However, we will show below 
that the stretching of redshift scale along the LoS will counteract this apparent
squashing producing a structure similar to a FoG.
In order to stay within the linear regime, we ensure not to reach the turnaround velocity 
by keeping  $g(\gamma,\beta,\mu{(\bm r)})$  
positive in the polar region.  In that case $\beta$ is limited from 0 to an upper limit which is a function of $\gamma$, and equals 2/3 when $\gamma = 1.8$. The   $\beta =   0$ case gives the no gravity one in which $\xi^{(s)}({\bm r}) = \xi^{(r)}({\bm r})$. 

We now remark that
$
 \mu{(\bm r)} = r_{||}/|{\bm r}| 
$ (see eq.~\ref{eq:dpmn31}).
But in some works { \citep[e.g.,][]{ms96, 2006MNRAS.368...85T, lc14}}  
it has been approximated as $\mu{(\bm s)} = s_{||}/|{\bm s}|$ or as
$
 \mu{(\bm{cs})} = c_{||} s_{||}/\sqrt{ c_{\bot}^2 s_{\bot}^2 + c_{||}^2 s_{||}^2 },
$ 
{ or even as $ r_{||} = s_{||}$}.
We { have referred  }   to this as the $\mu{(\bm s)}$ approximation.  In principle, given that $\mu$ is a scalar 
function, either form should be acceptable as long as the ${\bm s}$ and ${\bm r}$ vectors
refer to the same point.  However, we remark that $r_{||}$ differs from
$c_{||} s_{||}$ (see eq.~\ref{eq:dpmn18}), and that it is usually unknown, since 
in order to obtain it from $s_{||}$, the infall velocity field must be known. So these approximations should be carefully used. 

{
The result in our eq.~\eqref{eq:dpmn33}  has been derived for $r$-space, profiting on the difference between $r$- and $s$- spaces.  
Plotting this function directly in $r$-space as the independent variable, produces a squashed  structure for $\xi^{(s)}(\bm r)$. 
However, one wants to display the correlation function in $s$-space to compare with observations, not in $r$-space.  In order to do so, some authors perform the  $\mu(\bm s)$  
approximation  while others may plainly substitute $s$ for $r$ all the way in  eq.~\eqref{eq:dpmn33} and write   $\xi^{(s)}( \bm s) = g(\gamma, \beta, \mu( \bm s)) ~ \xi^{(r)}(s) $     to be able to display $\xi^{(s)}  $ in $s$-space. This is certainly wrong because $\bm s$ and $\bm r$ are not just independent names for position, and there exists a relation $\bm s(\bm r)$ between them that is not linear. 
Specifically, the parallel component is $s_{||} \sim r_{||} + v _{||} $ (eq.~\ref{eq:dpmn18}), with $v_{||}$ also a (yet unknown) function of position $\bm r$.  In the case of small disturbances we expect small velocities (below turnover) that result in a bi-univocal map $\bm s(\bm r)$ and its inverse.
So,  if we want to display the resulting $\xi^{(s)}$ in $\bm s$ space, one must proceed first to evaluate $\bm r= \bm r(\bm s)$ and then $\xi^{(s)}(\bm r)$ via eq.~\eqref{eq:dpmn33}, or in short $\xi^{(s)}(\bm r (\bm s ) )  =  g(\gamma, \beta, \mu( \bm r(\bm s ))) ~ \xi^{(r)}(r(\bm s ))  $.  We can therefore informally define $\xi^{(s)} (\bm s ) \equiv \xi^{(s)}(\bm r (\bm s ) ) $ and we claim that this is the correct way to evaluate the two-point correlation function on a grid in $s$-space. 

}

On the other hand, 
{ if the $\mu{(\bm s)}$} approximation is used one then obtains structures that are squashed in the LoS direction, and 
with a characteristic peanut-shaped geometry close to the polar axis 
\citep[see for example][]{hawkins2003}.  One concludes that this geometry fails to reproduce the structure
known as ``Finger of God" (FoG). The consequence is that other processes are called upon to account for it,
such as random motions arising in the virialized inner regions of clusters.
We next show below that by avoiding this approximation, it allows us to obtain a geometrical structure quite similar to the FoG feature.

\section{Projected correlation function}\label{sec:sec4}

In order to avoid the complications that redshift-space distortions introduced in the
correlation function, such as those
produced by gravitationally induced motions or virialized conditions, the projected
correlation function $w_{\bot}(r_{\bot})$ is frequently
preferred in the analysis.  This approach was first suggested in the analysis of CfA data by \citet{davis_peebles_83}, who mention that at
small redshift separations, peculiar velocities may cause $\xi(s)$ to differ from  $\xi(r)$.
To avoid this effect, they integrate $\xi(r)$ along the redshift difference to obtain the projected function $w_{\bot}(r_{\bot})$ on the plane of the sky.  Then, from it, they recuperate $\xi(r)$ inverting the problem by solving Abel's integral equation \citep{bt87} numerically.  See also \citet{pisiani2014} for other possibilities.  In the case where $\xi(r)$ is
a power-law, $w_{\bot}(r_{\bot})$ will be one as well, and the relation between them is analytical \citep[e.g.,][]{krumpe10}. 

We will show that the projected correlation function can be used to obtain 
the $r_{||}(s_{||})$ function that allows one to calculate $\mu{(\bm r)}$.
We start by noting that the projection on the plane of the sky may be performed 
either by using the $\xi^{(s)}$
function or its real space counterpart $\xi^{(r)}$.  Then we define the 
projected correlation functions as
\begin{equation}\label{eq:dpmn35}
 w^{(s)}_{\bot} (s_{\bot},s^*_{||}) = \int_0^{s^*_{||}}{ \xi^{(s)}({  \bm r} (s_{\bot},s_{||} ) ) ~ ds_{||} },
\end{equation} 
and
\begin{equation}\label{eq:dpmn36}
 w^{(r)}_{\bot} (r_{\bot},r^*_{||}) = \int_0^{r^*_{||}} { \xi^{(r)}(r_{\bot},r_{||}) ~ dr_{||} },
\end{equation} 
where  $\xi^{(s)}(   \bm r  (s_{\bot},s_{||} )   $, given by eq.~\eqref{eq:dpmn33}, may be understood as $\xi^{(s)}( \bm s )     $ as mentioned above.

The integral limits should go to infinity to get the total projected functions.  However,
one can project the correlation function up to a particular real space distance $r^*_{||}$.
Furthermore, if we assume that there exists a biunivocal function $s_{||}(r_{||})$, then we can find 
the corresponding $s^*_{||} = s_{||}(r^*_{||})$. 
{Boundary conditions are thus well defined \citep[e.g.,][]{Nock2010}. On the one hand slices in r-space (eq.~\ref{eq:dpmn36}) do not depend on the observers perspective, while on the other (eq.~\ref{eq:dpmn35})
the limit of the integral (boundary condition) becomes a function that is precisely going to be 
evaluated.} Carrying on,  due to number conservation
the projections in redshift- and real-space, multiplied by the corresponding area elements
that complete the volume where the number of pairs are counted, must be equal. Which leads to 
\begin{equation}\label{eq:dpmn37}
w^{(s)}_{\bot} (s_{\bot},s^*_{||}) ~ ds_{\bot}^2 =w^{(r)}_{\bot}(r_{\bot},r^*_{||}) ~ dr_{\bot}^2,
\end{equation} 
for all values of $r_{\bot}$ (or its corresponding $s_{\bot}$, see eq.~\ref{eq:dpmn9}). Inverting the
$s_{||}(r_{||})$ map and using  eqs.~\eqref{eq:dpmn35} to  \eqref{eq:dpmn37}, together with \eqref{eq:dpmn33} and \eqref{eq:dpmn9} we obtain
\begin{equation}\label{eq:dpmn38}
\int_0^{s^*_{{||} }} {  g(\gamma, \beta, \mu{(\bm r)})~ \xi^{(r)}(r_{\bot},r_{||}) ~ ds_{||} }  =   c_{\bot}^2 ~\int_0^{r_{||}(s^*_{{||} })} {    \xi^{(r)}(r_{\bot},r_{||}) ~ dr_{||}}.
\end{equation} 
Then, changing variables to $r_{||}$ in the left  ($ds_{||} = {\frac {ds_{||} } {  dr_{||} } } dr_{||} $), and noting that the equality holds 
for all values of $s^*_{{||} }$,   the integral signs can be omitted. Furthermore, 
using eqs.~\eqref{eq:dpmn15} and \eqref{eq:dpmn16} 
the equation simplifies to
\begin{equation}\label{eq:dpmn39}
  c_{||} ~ ds_{||} = {\frac { dr_{||} } {   g(\gamma, \beta, \mu (\bm r (r_{\bot},r_{||} ))) } }    ,
\end{equation} 
where the dependence $ \mu (\bm r (r_{\bot},r_{||} )) = r_{||} / \sqrt{r_{\bot}^2+r_{||}^2}$ has
been emphasized for clarity.  Equation \eqref{eq:dpmn39} completes the metric transformation between 
redshift- and real-spaces. As a consistency test, we note that in the limit of no gravitational
disturbance ($\beta = 0 $) we have $g(\gamma,\beta,\mu{(\bm r)}) = 1$ and eq.~\eqref{eq:dpmn8} is recovered.

\section{Resulting redshift-space and real-space relation}\label{sec:sec5}

\begin{figure}
\includegraphics[width=\columnwidth]{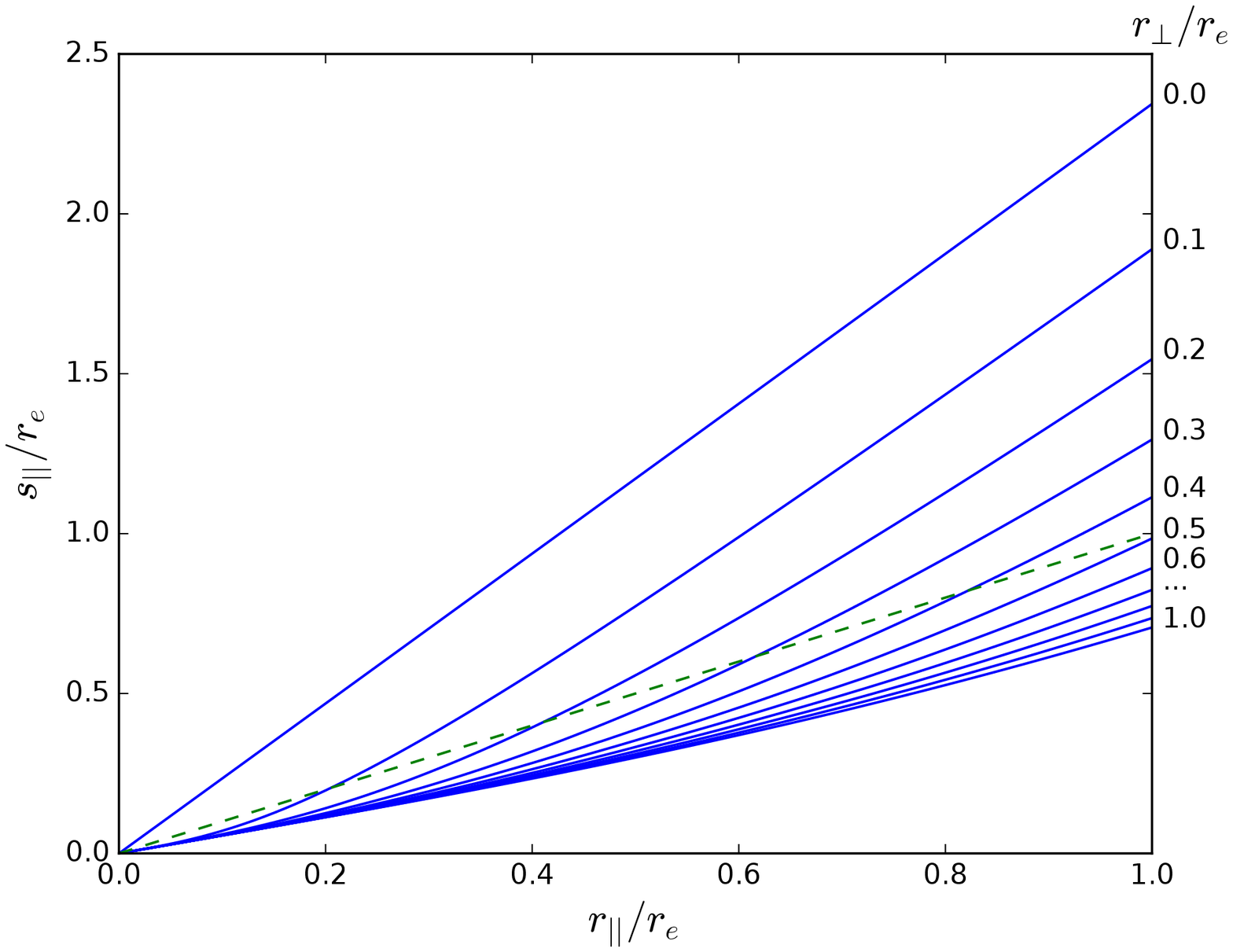}
\caption{ $s_{||}/r_e $ vs. $ r_{||}/r_e$  for  $r_{\bot}/r_e$ from 0 to 1 as indicated in the figure, for $\gamma = 1.8$, $\beta = 0.4$, and  $c_{||} =1$. For any value of the scaling parameter $r_e$. The dashed line indicates the identity $s_{||}= r_{||} $ for reference.}
\label{fig:rdes} 
\end{figure}

We integrate eq.~\eqref{eq:dpmn39} numerically using eq.~\eqref{eq:dpmn34}, to obtain the  $s_{||}(r_{||})$ function shown in Figure~\ref{fig:rdes}, for different values of $r_{\bot}/r_e$ indicated for each curve in the figure, where $r_e$ is an arbitrary scaling parameter, $\gamma = 1.8$, $\beta = 0.4$, and $c_{||} =1$. 
Note that the relation is
not linear.  If we compare to the identity line 
($s_{||} = r_{||} $) shown as a dashed line, we note that
sometimes the curves of constant $r_{\bot}$ lie above or below the identity line,
or even cross it. 

So, it can be noted that for on-axis
separations (where $r_{\bot} = 0$),   the spatial scale 
in redshift space is stretched, i.e. $s_{||} > r_{||} $,
effectively opposing the squashing effect obtained 
by the rough $\mu{(\bm s)}$ approximation. 
On the other hand, 
for $r_{\bot} \rightarrow 1 $ a squashed structure is seen (even more so that 
the one obtained by the $\mu{(\bm s)}$ approximation) that ultimately
converges to the limit $  s_{||} \rightarrow  r_{||} $ as we approach the plane of the sky ($ r_{||} = 0$).

\begin{figure}
\includegraphics[width=\columnwidth]{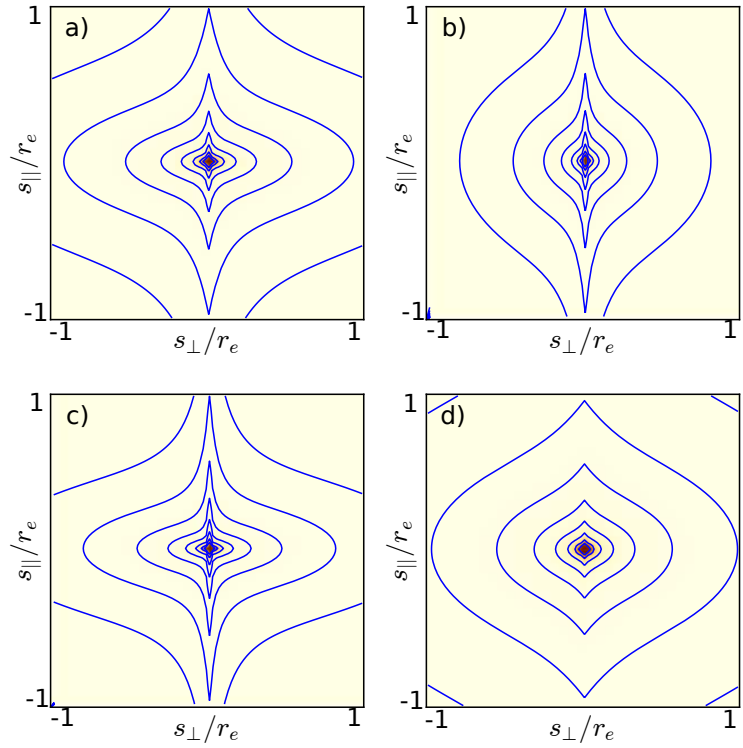}
\caption{ Redshift-space Two-Point-Correlation-Function (2PCF) $\xi^{(s)}(s_{\bot},s_{||})$ in logarithmically spaced contours at $e$ intervals  for any value of the scaling parameter $r_e$.  The parameter values are:
a) $\gamma = 1.8$, $\beta = 0.4$ and $AP =1$; 
b) $\gamma = 1.8$, $\beta = 0.4$ and $AP =2$; 
c) $\gamma = 1.8$, $\beta = 0.5$ and $AP =1$; 
d) $\gamma = 1.8$, $\beta = 0.2$ and $AP =1$;  }
\label{fig:figfinga}
\end{figure}

These geometrical distortions can be better appreciated by
their effect on the 2PCF presented in Fig.~\ref{fig:figfinga}. Here we start from a grid in 
$s-$space, and
transform to r-space using the integral relation (eq.~\ref{eq:dpmn39})  for the parallel component  and 
eq.~\eqref{eq:dpmn9} for the perpendicular one. From there, we calculate $\mu{(\bm r)}$ 
(eq. \ref{eq:dpmn31}), $g(\gamma, \beta, \mu{(\bm r)}) $ (eq. \ref{eq:dpmn34}), assuming that 
$\xi^{(r)}({\bm r}) =  (r/r_0)^{-\gamma}$; and finally, $\xi^{(s)}({\bm s})$ { (i.e. $\xi^{(s)}(\bm r ({\bm s}))$ ) }  from eq.~\eqref{eq:dpmn33}.  
The cosmological distortion is governed by the $c_{||}$ and $c_{\bot}$
parameters that depend on the Alcock-Paczy\'nski 
function $AP$ (see eq.~\ref{eq:dpmn12}). Its value
depends on the cosmological parameters 
${\bm \Omega} = (\Omega_m,\Omega_k,\Omega_\Lambda)$ and
increases with the redshift $z$ \citep[see figure 1 in][]{ap79}.

Figure~\ref{fig:figfinga}(a) shows the
case that corresponds to the parameters used for
Figure~\ref{fig:rdes}: $\gamma = 1.8$, $\beta = 0.4$ and $AP =1$,
where the geometrical distortions produced are evident, an elongation in the polar direction and a squashing in the equatorial direction. As can
be noted the polar elongation resembles the structure known as
FoG. 

In the other three figures, \ref{fig:figfinga}(b), \ref{fig:figfinga}(c) and \ref{fig:figfinga}(d),  we explore the effect of
cosmological and gravitational alterations. Figure~\ref{fig:figfinga}(b)
shows that the effect of increasing $AP$ is a geometrical distortion that
concentrates the structure towards the polar axis direction for
$AP=2$ that corresponds to $\Lambda$CDM cosmology at $z=2.6$.
In Figure~\ref{fig:figfinga}(c)  we explore the effect of changing the dimensionless growth-rate for visible mater $\beta$.  This gravitational effect
is to enhance the FoG structure as its value increases (recall that its limit value is 2/3). On the other hand, if $\beta$ decreases the structure becomes rounder and the FoG faints
accordingly as is shown in Fig. \ref{fig:figfinga}(d). By comparing figures \ref{fig:figfinga}(b) and \ref{fig:figfinga}(c)  relative to panel \ref{fig:figfinga}(a), we note that the same enhanced strength of the FoG feature is obtained in the small scale regions, but the large scale structure is quite different. This is because in the first case the distortion is cosmological while on the second it is gravitational.

Although it has not been the purpose of this paper, we may consider different values of the power-law index $\gamma$ and obtain figures similar to those shown in Fig.~\ref{fig:figfinga}.  In some cases they might even resemble some of the cases depicted here.  It turns out that lower values may accommodate rounder 2PCFs at mid scales, while a steeper $\gamma$ may also concentrate the structure towards the LoS.  Note however, that it is easy to discern those cases by a simple projection on the plane of the sky, as depicted through section~\ref{sec:sec4}. This is because that projection will erase redshift distortions, both gravitational ($\beta$) and cosmological ($AP$) while preserving the radial structure $\gamma$.

As we have indicated, a rounder 2PCF at mid spatial scales is favoured by some works that use the $\mu(s)$ approximation. 
As can be seen in Fig.~\ref{fig:figfinga}(b), rounder figures can be obtained with lower
values of $\beta$.  We have estimated that a $\beta = 0.25 $ produces a 2PCF which is equally
squashed to that obtained by the $\mu(s)$ approximation for the case $\beta = 0.4 $ for most
points in the s-space plane, those with $s_{\bot} > s_{||}$. An increase in the $AP$ parameter may also contribute to alleviate the situation.

Another possibility, that was not intended to be covered here, is the case of
a more realistic 2PCF $\xi^{(r)}({ r} )$ as the ones inferred from baryon acoustic oscillations (BAOs) \citep[e.g.][]{slosar2013}  or those  obtained by the CAMB code \citep{seljak_zaldarriaga_96}.  
In order to apply the results of this paper to such cases, one could try breaking the inferred $\xi^{(r)}({ r})$ profile in a series of power-laws and then apply eq.~\eqref{eq:dpmn39} to each section. If this is not possible, then we would have to give up eqs.~\eqref{eq:dpmn33}  and \eqref{eq:dpmn34}  as a
way of simplifying $\xi^{(s)}({\bm r})$.  However, the projections in the plane of the
sky, i.e. eqs.~\eqref{eq:dpmn35}  and \eqref{eq:dpmn36}   are still valid, and instead of using eq.~\eqref{eq:dpmn33} 
to simplify, we would have to go back to the expansion of $\xi^{(s)}({\bm r})$ in multipoles eq.~\eqref{eq:dpmn30}.  In that case one would end up with the following equation:
\begin{equation}\label{eq:dpmn40}
 c_{||} ~ ds_{||} = {\frac { \xi^{(r)}(r)  }  { \sum_{l=0,2,4} \xi_l(r) ~ P_l(\mu{(\bf r)})   } }  dr_{||}  
 \end{equation}
instead of eq.~\eqref{eq:dpmn39}.  And we would have to find a way to estimate the multipole moments $\xi_l(r)$.  Another possibility
is to
leave $\xi^{(s)}(r)$ in the denominator. Considering these possibilities seems like an interesting task for future works, but it is beyond the scope of this paper.

We conclude that a whole range of possibilities in shape and
strength of the FoG structure and the squashing of the
equatorial zone can be obtained by tuning the parameters
$\gamma$,  $\beta$,  and $AP$. This may provide a path towards
solving the usual degeneracy problem between cosmological and 
gravitational distortions,
that can still be seen at a level of
10\% in 1$\sigma$ correlated variations in recent
work \citep[e.g.][]{Satpathy2017}.

\section{Conclusions}\label{sec:sec6}

We emphasize the importance of distinguishing three spaces in cluster and large scale structure studies: the observable
redshift-space ${\bm \sigma}$, the physical redshift-space ${\bm s}$, and the
real-space ${\bm r}$. The transformation between ${\bm \sigma}$ and ${\bm s}$ 
is isotropic dilation that introduces a scale factor dependent on the 
cosmology. 

On the other hand, the transformation between ${\bm s}$ and 
${\bm r}$ goes through a unitary Jacobian independent of redshift, and only
distorts the space by factors related to the Alcock-Paczy\'nski $AP$ function
(c.f., eqs.~\ref{eq:dpmn15} and \ref{eq:dpmn16}).

Furthermore, when we introduce peculiar non-relativistic velocities 
in this scheme, we demonstrate that the same relation between observable 
and physical redshift-spaces 
${\bm s} = K({\bm \Omega},z) ~ {\bm \sigma}$ is kept. In the analysis of the 2PCF in the physical redshift-space ${\bm s}$, we recover
the \citet{kaiser87} effect independent of redshift in Fourier space, and 
\citet{hamilton92} results in configuration space.  

We remark,  that
there appears a dependence with $\mu$ in real-space  
($ \mu{(\bm r)} = r_{||}/|{\bm r}|$), and that it has been a common practice to
approximate it  from redshift-space coordinates 
as either $ \mu{(\bm s)} = s_{||}/|{\bm s}|$ or  
$\mu{(\bm {cs})} = c_{||} s_{||}/\sqrt{ c_{\bot}^2 s_{\bot}^2 + c_{||}^2 s_{||}^2 }$ or 
$r_{||} = s_{||}$, 
sometimes called the``distant observer approximation", or simply to substitute $\bm s$ for $\bm r$ in the equations. To avoid further confusion we have called this  the  $\mu{(\bm s)}$ approximation in any of its forms.
We argued that this wrong
assumption produces either a squashed or a peanut-shaped geometry close to the
LoS axis, for the 2PCF in redshift-space.

Since $r_{||}$ is usually unknown, we proposed a method to
derive it from $s_{||}$ using number conservation in the
projected correlation function in both real- and 
redshift-spaces. This led to a closed form 
eq.\,\eqref{eq:dpmn39} for the case where the real 2PCF can be
approximated by a power-law. From this, we solved for
$\mu{(\bm r)}$ in real-space, and showed that a different view
of the redshift-space 2PCF emerges. The main result is that
the redshift-space 2PCF presents a distortion in the LoS
direction which looks quite similar to the 
ubiquitous FoG. This is due to a strong anisotropy that 
arises purely from
linear theory   and   produces   a stretching of the scale as
one moves into the on-axis LoS direction.   
Moving away from the LoS the structures appear somewhat more squashed 
than those obtained by the $\mu(s)$ approximation for 
equivalent values of $\beta$. The implications of this remains an open question.

The development presented here produces structures that 
qualitatively  reproduce the observed features of the 2PCF of
galaxies and quasars large scale structure.  A squashing
distortion in the equatorial region is attributed to a 
mixture of cosmological  and gravitational effects. And the
FoG feature that is usually attributed to other causes, is
instead ascribed to the same gravitational effects derived
from linear theory. 

We conclude that a whole range of possibilities in shape
and strength of the FoG structure, and the squashing of the
equatorial zone, can be obtained by tuning the parameters
{$\gamma$}, {$\beta$},  and { $AP$}. This provides a path towards
solving the usual  degeneracy problem between cosmological and gravitational distortions. In a future paper (Salas \& Cruz-Gonz\'alez in preparation) we apply these results to the galaxies and quasar data obtained by current large scale surveys.

\bigskip

Acknowledgements. I.C.G. acknowledges support from DGAPA-UNAM (Mexico) grant IN113417.

\bibliography{cosmologia}


\end{document}